\documentclass[10pt,twocolumn,letterpaper]{article}

\usepackage{cvpr}              

\usepackage[dvipsnames]{xcolor}

\usepackage{multirow}
\usepackage{booktabs}
\usepackage{pbalance}

\definecolor{cvprblue}{rgb}{0.21,0.49,0.74}
\usepackage[pagebackref,breaklinks,colorlinks,citecolor=cvprblue]{hyperref}

\def\paperID{5} 
\def\confName{CVPR}
\def\confYear{2024}

\title{ChatVTG: Video Temporal Grounding via Chat with \\ Video Dialogue Large Language Models}

\author{Mengxue Qu$^{1,2}$\thanks{Work done during an internship at JD Explore Academy. } \quad
Xiaodong Chen $^3$ \quad
Wu Liu $^3$ \quad
Alicia Li $^{4}$ \quad
Yao Zhao$^{1,2}$\\
\\
$^1$Institute of Information Science, Beijing Jiaotong University \\
$^2$Beijing Key Laboratory of Advanced Information Science and Network Technology \\
$^3$University of Science and Technology of China   \quad
$^4$Horace Mann School \\
}

\begin{document}
\maketitle
\begin{abstract}
Video Temporal Grounding (VTG) aims to ground specific segments within an untrimmed video corresponding to the given natural language query.  
Existing VTG methods largely depend on supervised learning and extensive annotated data, which is labor-intensive and prone to human biases.  
To address these challenges, we present ChatVTG, a novel approach that utilizes Video Dialogue Large Language Models (LLMs) for zero-shot video temporal grounding. 
Our ChatVTG leverages Video Dialogue LLMs to generate multi-granularity segment captions and matches these captions with the given query for coarse temporal grounding, circumventing the need for paired annotation data.
Furthermore, to obtain more precise temporal grounding results, we employ moment refinement for fine-grained caption proposals.  
Extensive experiments on three mainstream VTG datasets, including Charades-STA, ActivityNet-Captions, and TACoS, demonstrate the effectiveness of ChatVTG.
Our ChatVTG surpasses the performance of current zero-shot methods.

\end{abstract}    
\section{Introduction}
\label{sec:intro}

With the rapid development of short video platforms, videos have become the primary form of documenting and sharing life experiences for people worldwide. Unlike static image data, videos encapsulate dynamic information, providing a more vivid and immersive way to showcase stories and memories. Understanding the content of videos is crucial as it enables meaningful interactions based on the video context~\cite{MeViS, MOSE:conf/iccv/DingLH0TB23, Partlevel:conf/iscas/ChenLL0W0022, ying2023ctvis}. Today's Video Understanding Large Language Models (LLMs)~\cite{VideoChat2:journals/corr/abs-2311-17005, VideoLLaVA:journals/corr/abs-2311-10122,VideoChatGPT:journals/corr/abs-2306-05424}, have demonstrated remarkable capabilities in engaging in dialogues based on video content, enhancing the model's ability to explore and comprehend the information conveyed within videos.

\begin{figure}[t]
  \centering
    \includegraphics[width=0.8\linewidth]{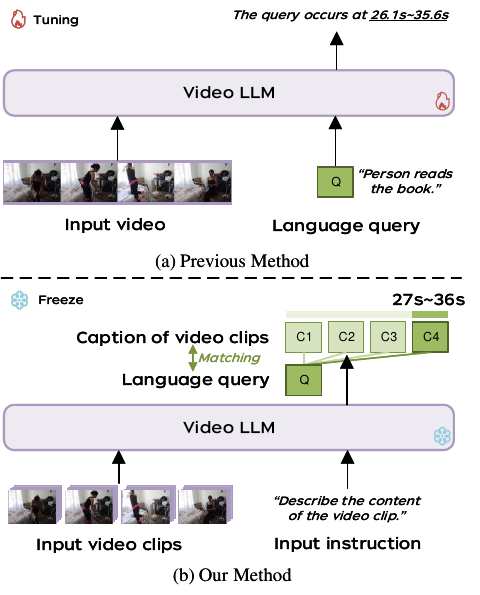}
  \caption{Comparison between previous methods based on Video LLMs and our approach: Previous methods require fully supervised training of the Video LLM, whereas our method does not require training and can provide temporal grounding zero-shot.
  }
  \vspace{-4mm}
  \label{fig:intro}
\end{figure}

As video content expands, efficiently retrieving specific segments from extensive video libraries remains with the rapid development of short video platforms, videos have become the primary form of documenting and sharing life experiences for people worldwide. Unlike static image data, videos encapsulate dynamic information, providing a more vivid and immersive way to showcase stories and memories. Understanding the content of videos is crucial as it enables meaningful interactions based on the video context. Today's Video Understanding Large Language Models (LLMs)~\cite{VideoChat2:journals/corr/abs-2311-17005, VideoLLaVA:journals/corr/abs-2311-10122,VideoChatGPT:journals/corr/abs-2306-05424}, have demonstrated remarkable capabilities in engaging in dialogues based on video content, enhancing the model's ability to explore and comprehend the information conveyed within videos. crucial but challenging. The task of Video Temporal Grounding (VTG) has been developed to address this by aiming to accurately locate continuous time windows within videos using natural language descriptions, which is essential for enhancing the accessibility and usability of video content amidst improved yet still imperfect comprehension capabilities.
Most existing VTG methods~\cite{2DTAN:conf/aaai/ZhangPFL20, MMN:conf/aaai/00010WLW22, VDI:conf/cvpr/LuoHGJL23, VTimeLLM:journals/corr/abs-2311-18445} heavily rely on supervised training with paired annotation data, which requires extensive effort for precise time labeling.
Additionally, the annotation process is further complicated by subjective biases in establishing precise temporal boundaries.

Compared to temporal grounding datasets, text-dialogue video data is more abundant and simpler to gather from the internet, with the potential for easier automated annotation. This data is often employed for tasks like video captioning and video question answering, where Video Language Models (LLMs) trained on it demonstrate notable conversational skills. Therefore, we contemplate \textit{whether it's feasible to leverage Video Dialogue LLMs to zero-shot retrieve the corresponding video segments for a given query.}

In this paper, we introduce an innovative approach that leverages Video Dialogue LLMs for video temporal grounding without the need for additional training data or paired annotations. As shown in~\cref{fig:intro}, different from existing methods that utilize VTG data to train Video LLMs for temporal grounding capabilities, our approach combines captions and matching to find the moment without training the Video LLM. Specifically, we start by dividing the video into multiple coarse segments and guide the Video LLM to provide captions for each segment through multi-granularity instructions. Then, we match these segments with the query provided by the user to identify coarse moments. To further refine the temporal boundaries, we conduct moment refinement by generating additional moment proposals using a sliding window approach. We then generate captions for proposals that overlap significantly with the coarse moments and further match them with the query to determine the final moment.
This approach allows for seamless integration with existing Video LLMs, eliminating the need for additional training or specialized datasets.

To validate the effectiveness of our proposed framework, we conduct experiments on three commonly used VTG datasets, including Charades-STA~\cite{CharadesSTA:conf/iccv/GaoSYN17}, ActivityNet-Captions~\cite{ActivityNetCaptions:conf/iccv/KrishnaHRFN17}, TACoS~\cite{TACoS:journals/tacl/RegneriRWTSP13}. Our ChatVTG outperforms existing zero-shot methods.

\section{Related Work}
\label{sec:related work}

\subsection{Video Temporal Grounding}
Video Temporal Grounding (VTG), also known as Video Temporal Localization (VTL) or Video Moment Retrieval (VMR), aims to align or ground a natural language query (often a sentence or phrase) to the corresponding moment or segment within an untrimmed video.
Based on the difference in experimental setups and types of supervision during training, we categorize the related studies into (a) fully supervised VTG, (b) weakly supervised VTG, (c) unsupervised VTG, and (d) zero-shot (a.k.a language-free) VTG.

\noindent{\textbf{Fully supervised VTG.}}
For this approach, previous methods~\cite{2DTAN:conf/aaai/ZhangPFL20, MMN:conf/aaai/00010WLW22, VDI:conf/cvpr/LuoHGJL23, UniVTG:conf/iccv/LinZCPGWYS23} first extract visual features and textual features from pre-trained models~\cite{bert:journals/corr/abs-1910-01108, 3DCNN:journals/access/JingHL19, CLIP:conf/icml/RadfordKHRGASAM21}.
After that, a multi-model model is trained on a large amount of labeled data, where each video is paired with a corresponding textual description. 
The model learns to predict the grounding results based on the video input. 
For instance, VDI~\cite{VDI:conf/cvpr/LuoHGJL23} generate textual and visual features from the Contrastive Language-Image Pretraining (CLIP) model~\cite{CLIP:conf/icml/RadfordKHRGASAM21} and propose a generic visual-dynamic injection model for capturing the temporal video changes and aligning them with the corresponding phrases (e.g. verb).
While fully-supervised models can achieve high performance on specific datasets, they are the most data-intensive approaches, as they require a large dataset with accurately labeled video-text pairs.

\noindent{\textbf{Weakly supervised VTG.}}
To mitigate the fine-grained labeling problem of fully supervised VTG, weakly-supervised methods~\cite{VCA:conf/mm/WangCJ21, CNM:conf/aaai/ZhengHCL22, CPL:conf/cvpr/ZhengHCPL22, Huang:conf/cvpr/HuangYS23} are proposed to learn the segment-text alignment without marking the precious starting and ending time.
During the training stage, only the video-level video-sentence matching is used for supervision.
To predict the precision temporal localization under this setting, some methods~\cite{Weak1:conf/iccv/HuangLGJ21, Weak2:conf/cvpr/MithunPR19} utilize the multi-instance learning (MIL) technique. 
In detail, these methods form positive and negative video-level video-language pairs from the original dataset. 
Then they amplify the matching scores for the correct pairs and diminish those of the incorrect ones. 
However, an untrimmed video contains multiple events and can not correspond exactly with the sentence describing a specific event at the video level.
Therefore, some methods~\cite{CPL:conf/cvpr/ZhengHCPL22, Huang:conf/cvpr/HuangYS23} apply the contrastive proposal learning or cycle self-training framework to generate positive and negative proposal at the frame level to optimize the model training under weak supervision.

\noindent{\textbf{Unsupervised VTG.}}
Compared with Weakly supervised VTG, the unsupervised methods~\cite{PSVL:conf/iccv/NamAKHC21, Gao:journals/tcsv/GaoX22, DSCNet:conf/aaai/LiuQWDZCXZ22, PZVMR:conf/mm/WangWLY22} further removes the paired video-level annotated sentences.
Due to the lack of supervision information, these methods optimize the model with pseudo queries or semantic features of the entire quire collection.
For example, PSVL~\cite{PSVL:conf/iccv/NamAKHC21} first generates the simplified pseudo-quire from the query set of unpaired annotated annotations and then simulates the multi-modal model with the pseudo-quire and video segment from the temporal video proposal module.
Unlike PSVL, DSCNet~\cite{DSCNet:conf/aaai/LiuQWDZCXZ22} propose a linguistic semantic mining module to extract the implicit semantic feature from the unpaired entire query set.
These linguistic semantic features guide the composition of the video activities, filtering out the redundant backgrounds and grounding the corresponding events. 
To be noted, some studies~\cite{PSVL:conf/iccv/NamAKHC21, PZVMR:conf/mm/WangWLY22} treat these methods as the zero-shot task. 
However, these methods still have access to unpaired training data, for which we categorize them as unsupervised tasks.

\noindent{\textbf{Zero-shot VTG.}}
With a strict zero-shot setting for real-world applications, zero-shot VTG~\cite{UniVTG:conf/iccv/LinZCPGWYS23}, sometimes referred to as language-free VTG, involves completing the video segment grounding corresponding to the given queries without relying on the training dataset.
Therefore, these models have no access to training data even unpaired data, and rely entirely on their pre-existing knowledge and generalization capabilities to perform the grounding task.
In practical terms, a zero-shot VTG model should typically be pre-trained on a diverse and extensive dataset without the limitation of tasks that enable it to learn a wide array of visual and linguistic concepts. 
This pre-training allows the model to develop an understanding of how language can describe video content. 
Then, when presented with the special zero-shot VTG task, the model utilizes this multi-modal comprehension to make inferences about where the described events are likely to occur in a video, even though it has never seen those specific quires during training.
For instance, LLM4VG~\cite{LLM4VG:journals/corr/abs-2312-14206} evaluates the performance of existing LLMs under the zero-shot setting.
UniVTG~\cite{UniVTG:conf/iccv/LinZCPGWYS23} propose a multi-modal and multi-task learning pipeline to pre-training the unified large-scale multi-modal model with dozens of unlimited datasets. 
In the inference stage, UniVTG is applied for the video temporal grounding task without fine-tuning or training on specific datasets.

\subsection{Multi-model Large Language Model}
Multi-modal Large Language Models (MLLM)~\cite{VideoChatGPT:journals/corr/abs-2306-05424, VideoChat2:journals/corr/abs-2311-17005}, like other Large Language Models (LLM), are trained on diverse and extensive multi-model datasets. 
VideoLLaVa~\cite{VideoLLaVA:journals/corr/abs-2311-10122} perform visual reasoning capabilities on both images and videos simultaneously with the binding of unified visual representations to the language feature space. 
Video-ChatGPT~\cite{VideoChatGPT:journals/corr/abs-2306-05424} is a video conversation model designed for generating meaningful conversations about the given videos. This model combines the capabilities of LLMs with a pre-trained visual encoder adapted for spatio-temporal video representation.
VideoChat2~\cite{VideoChat2:journals/corr/abs-2311-17005} is a robust video MLLM baseline for multi-model reasoning task. This model is progressive multi-modal trained with diverse instruction-tuning data to achieve temporal understanding in dynamic video tasks.
The multi-modal comprehension of these models allows them to capture complex patterns and perform a wide variety of tasks.
For example, they could potentially describe the content of a given image or video in text and have conversations with the user.

LLaViLo~\cite{LLaViLo:conf/iccvw/MaZFFBWHLS23} exploits the capabilities of these MLLM in video understanding and designs a specialized adapter for the VTG task.
With the joint training of a set prediction objective and a captioning objective, LLaViLo achieves significant performance improvement on the fully supervised VTG.
VTimeLLM~\cite{VTimeLLM:journals/corr/abs-2311-18445} and PG-Video-LLaVA~\cite{PGVideoLLaVA:journals/corr/abs-2311-13435} leverage the powerful conversational capabilities of Vicuna~\cite{Vicuna:zheng2023judging} and LLaVA~\cite{llava:conf/nips/LiuLWL23a} to treat VTG tasks as Q\&A tasks, achieving robust fully-supervised performance with fine-tuning.
However, the reliance on labeled VTG datasets reduces these models' ability to generalize to real-world scenarios that differ from the training data.
To overcome these shortcomings, this paper attempts to accomplish the zero-shot VTG tasks utilizing the video comprehension capabilities of pre-trained MLLMs.
\section{Method}
\begin{figure*}[!t]
  \centering
    \includegraphics[width=0.9\linewidth]{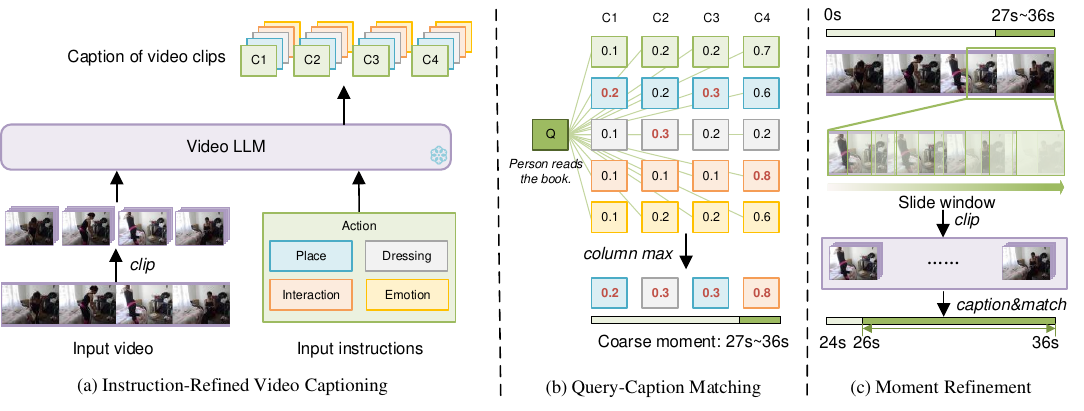}
  \caption{Pipeline of ChatVTG. Our method pipeline primarily consists of three components: (a) Instruction-Refined Video Captioning; (b) Query-Caption Matching; and (c) Moment Refinement. Best viewed in color.
  }
  \vspace{-4mm}
  \label{fig:arch}
\end{figure*}

In this section, we commence by elucidating the task definition of Video Temporal Grounding (VTG) and provide an overview of existing generic frameworks utilized for Video Large Multimodal Models (LMMs). Following this, we introduce our novel approach, which capitalizes on Video LMMs without the need for training, and proceed to expound upon its details.

\subsection{Task Definition}
The goal of the VTG task is to accurately pinpoint video segments within untrimmed videos that correspond to the description provided by a given query description. This can be formally expressed as:
{
\small
\begin{equation}
[ts, te] = F(V, Q),
\end{equation}
}

where \(ts\) and \(te\) represent the start and end timestamps of the located video segment, respectively. Here, \(F(\cdot)\) denotes the VTG model, \(V\) denotes the input untrimmed video, and \(Q\) represents the query description.

In this study, our objective is to directly utilize Video LMMs to achieve temporal grounding and query matching for the VTG task. These Video LMMs are trained solely on natural dialogues, enabling proficient performance in tasks such as video captioning and video question answering, contributing to a comprehensive understanding of video content. However, they lack training in temporal comprehension, thus unable to directly output moments through dialogue. This setup is defined as zero-shot, wherein the model is not trained on specific VTG datasets yet is expected to perform VTG tasks. In previous zero-shot approaches, some level of training was still required~\cite{PSVL:conf/iccv/NamAKHC21, PZVMR:conf/mm/WangWLY22}. In contrast, our approach adopts a completely offline, training-free method, obtaining temporal outputs through dialogue with Video LMMs.

\subsection{Review of Video LMMs}
With the advancement of LMMs, numerous works have focused on training to establish Video Instruction datasets, enabling LMMs to comprehend video content and engage in dialogue based on it. In this section, we briefly review the common network architectures and training pipeline of Video LMMs.

\noindent {\textbf{Model Architecture}}
The model architecture of Video LMMs typically involves several components. Initially, visual encoders extract features from raw visual signals. These features are then linearly mapped to the textual feature space, creating a unified visual representation. Subsequently, shared projection layers encode this representation before integration with textual queries, which are then inputted into a Large Language Model (LLM) to generate responses. Alignment between the video and language spaces converges into a unified visual feature space, bridging the gap between different visual signals and preparing them for input into the large language model to generate dialogue. 

\noindent {\textbf{Training Pipeline}}
The training pipeline of Video LMMs consists of two main stages: understanding training and instruction tuning. During the basic understanding training phase, the model is trained to interpret visual signals using an extensive dataset of image/video-text pairs. Each visual signal corresponds to a single round of conversation data \((Xq, Xa)\), where \(Xq\) represents the input query and \(Xa\) is the ground truth response. The training objective involves the original auto-regressive loss, enabling the model to learn the basic ability to interpret visual information. Parameters unrelated to vision interpretation are frozen during this process. In the instruction tuning phase, the model is tasked with providing responses based on different instructions, which typically involve more complex visual comprehension tasks beyond simply describing visual signals. 

Upon reviewing Video LLMs, it becomes apparent that traditional methods for completing the VTG task require training Video LLMs with paired \textit{video-query-moment} data, a process attempted by some prior studies~\cite{VTimeLLM:journals/corr/abs-2311-18445, PGVideoLLaVA:journals/corr/abs-2311-13435}. However, this approach is characterized by its time-consuming and labor-intensive nature. In this paper, we propose a straightforward approach to address the VTG task, utilizing existing Video Dialogue Large Models. Our strategy involves generating captions and subsequently matching them with queries, aiming to efficiently achieve VTG.

\subsection{ChatVTG}
As illustrated in~\cref{fig:arch}, our method pipeline primarily consists of three components: (a) Instruction-Refined Video Captioning; (b) Query-Caption Matching; and (c) Moment Refinement. The specific details of each component are elaborated below.

\subsubsection{Instruction-Refined Video Captioning}
As shown in~\cref{fig:arch} (a), given the untrimmed video for query, to facilitate effective temporal grounding, we initiate the process by segmenting the video \(V\) into \(m\) equal-length video clips, thus obtaining segments that are more manageable for analysis,
{
\small
\begin{equation}
    V \rightarrow \{S_1, S_2, ..., S_m\}
\end{equation}
}
Where \(S_i\) denotes the \(i\)th video clip, and each clip has a time duration \(T\).
Subsequently, we feed each segmented video clip \(S_i\) into the Video LLM, leveraging its capabilities to comprehend and interpret visual content. By providing the Video LLM with a specific instruction tailored to the task, we guide the model to generate captions for each video segment,
{
\small
\begin{equation}\label{eq:videollm}
     \text{C}_i  = \text{Video LLM}(S_i, I)
\end{equation}
}
Where \(C_i\)  represents the generated caption for the \(i\)-th video clip, \(I\) represents the specific instruction tailored to the task provided to the Video LLM.

Due to the diverse nature of queries in the VTG task, which may describe events from various perspectives, we have established multi-granularity instructions to ensure a comprehensive understanding of video content and provide detailed descriptions. This aims to avoid mismatches between queries and captions resulting from overlooking important details. These instructions cover different aspects of the video, facilitating more nuanced guidance to ensure that the generated captions align with the video content. Specifically, we include the following five multi-granularity instructions, including ``action", ``place", ``dressing". ``emotion" and ``interaction":

{
\footnotesize
\begin{align}
I_{\text{action}} &= \text{``Describe the action of the person in the video."} \nonumber\\
I_{\text{place}} &= \text{``Where does this video take place?"} \nonumber\\
I_{\text{dressing}} &= \text{``What are the people in the video wearing?"} \label{eq:5}\\  
I_{\text{emotion}} &= \text{``illustrate the person's emotion or facial expression."}\nonumber\\
I_{\text{interact}} &= \text{``Describe the interaction of person and other people or things."}\nonumber
\end{align}
}

Through these multi-granularity instructions, we can achieve a more comprehensive understanding of various aspects of the video content, thereby providing more accurate and detailed descriptions for video caption generation, The process can be represented as:

{
\small
\begin{align}
     I_{mult} &= \{I_{\text{action}}, I_{\text{place}}, I_{\text{dressing}}, I_{\text{expression}}, I_{\text{interaction}}\} \\
    C_{i} &= \text{{VideoLLM}}(S_{i}, I_{mult})  \\
   &=  \{C_{i,\text{action}}, C_{i,\text{place}}, C_{i,\text{dressing}}, C_{i,\text{expression}}, C_{i,\text{interaction}}\} \nonumber
\end{align}
}

This formulation encapsulates the essence of our approach, wherein the Video LLM is employed to generate descriptive captions for each segmented video clip based on the provided instruction.

\subsubsection{Query-Caption Matching}
\label{sec:query-caption matching}
After obtaining multi-dimensional captions for each video clip, we proceed to match these captions with the query. Specifically, we employ SentenceBERT~\cite{SentenceBERT:conf/emnlp/ReimersG19} to encode both the captions and the query, 
{
\small
\begin{align}
    f_{Q} &= SentenceBERT(Q), \\
    f_{C_{i}} &= SentenceBERT(C_{i})
\end{align}
}
We calculate their cosine similarity,
{
\small
\begin{equation}
   \text{{Cosine Similarity}}(f_{C_{i}}, f_{Q}) = \frac{{f_{C_{i}} \cdot f_{Q}}}{{\|f_{C_{i}}\| \|f_{Q}\|}} 
\end{equation}
}
where \(f_{C_{i}}\) represents the caption feature for the \(i^{th}\) video clip, and \(f_{Q}\) represents the query feature. 
Then we get a cosine similarity score matrix. 
Let \( S \) denote the similarity matrix, where rows represent different granularity captions, and columns represent different video clips. Each entry in the matrix represents the cosine similarity score calculated between the caption and query for a specific video clip. The similarity matrix \( S \) can be represented as:

{
\small
\begin{equation}\label{matrix}
S = \begin{bmatrix}
s_{11} & s_{12} & \cdots & s_{1m} \\
s_{21} & s_{22} & \cdots & s_{2m} \\
\vdots & \vdots & \ddots & \vdots \\
s_{n1} & s_{n2} & \cdots & s_{nm}
\end{bmatrix}
\end{equation}
}
where \( n \) is the number of different granularity captions and \( m \) is the number of video clips. Each \( s_{ij} \) represents the cosine similarity score between the caption generated from the \( i^{th} \) granularity instruction and the query for the \( j^{th} \) video clip.

To ensure fair comparison across different granularity levels and comprehensively consider the maximum potential similarity between each granularity caption and the query, we employ Column-wise Maximum Normalization. This method selects the maximum similarity score \(s_{\text{max}}^{(j)}\) for each column \(j\) (representing a video clip) across all granularity levels from the similarity matrix \(S\), denoted as \(S_{\cdot j}\), and then normalizes these scores. By doing so, it accounts for the highest potential match between each video clip and the query while allowing for fair comparison among different granularity levels. This normalization process scales the similarity scores proportionally, ensuring that the maximum similarity score for each column becomes $1.0$, while preserving the relative relationships among the scores within each column. This process is illustrated in~\cref{fig:arch} (b).

After obtaining the normalized scores, we need to select the best clip or combinations of clips with closely related scores to determine the final moment window. This is achieved by setting a threshold based on the cosine similarity scores of the video clips. We select the longest consecutive combination of clips with scores exceeding the threshold as the final moment window.

\subsubsection{Moment Refinement}
Given the segmentation of the video into non-overlapping \( m \) clips in the previous step, it's essential to acknowledge that the initial moment boundaries might lack the precision needed for accurate temporal grounding. To address this, we adopt a strategy to refine these boundaries and enhance the fidelity of the moment detection process as in~\cref{fig:arch} (c).

Firstly, we segment the video using a sliding window approach into smaller segments. For a video of length \(L\), we use ``wide" to represent the window size and ``step" to represent the step size, where \(wide < L\) and \(step < wide\). The number of windows is denoted by \(k = \left\lfloor \frac{{L - \text{wide}}}{\text{step}} \right\rfloor + 1\). The start time of the \(i\)th window is \(t_i = (i-1) \times \text{step}\), and the end time is \(t_i + \text{wide}\). Therefore, the \(i\)th window corresponds to the video segment \(S_i\), where \(i = 1, 2, ..., k\).

Subsequently, we select the window combinations that have an intersection over union (IoU) greater than 0.7 with the coarse moment proposals obtained initially. This significant overlap between the selected windows and the proposed segments indicates their importance in capturing essential content from the video. Next, we input these selected window combinations into the VideoLLM for caption generation as in~\cref{eq:videollm}. This approach allows us to capture important details. 

Query-caption matching proposed in~\cref{sec:query-caption matching} is then performed again to determine the final moment. In summary, with moment refinement, we are able to obtain more precise moment boundaries.

\section{Experiments}
In this section, we first introduce benchmark datasets, evaluation metrics, and implementation. 
After that, we compare the quantitative results of our ChatVTG with existing methods. 
At last, we provide ablation studies of each module.

\begin{table*}
\centering
\scalebox{0.8}{
\setlength{\tabcolsep}{3mm}{
\begin{tabular}{c|c|c|cccc|cccc} 
\toprule
\multirow{2}{*}{Method} & \multirow{2}{*}{Year} & \multirow{2}{*}{Setup} & \multicolumn{4}{c|}{Charades-STA}                                 & \multicolumn{4}{c}{ActivityNet-Captions}                          \\ 
\cline{4-11}
                        &                       &                        & R@0.3          & R@0.5          & R@0.7          & mIoU           & R@0.3          & R@0.5          & R\%0.7        & mIoU            \\ 
\hline
2D-TAN~\cite{2DTAN:conf/aaai/ZhangPFL20}                  & ICCV'20               & \multirow{4}{*}{FS}    & 57.31          & 45.75          & 27.88          & 41.05          & 60.43          & 43.41          & 25.04         & 42.45           \\
MMN~\cite{MMN:conf/aaai/00010WLW22}                     & AAAI'22               &                        & 65.43          & 53.25          & 31.52          & 46.46          & 64.48          & 48.24          & 29.35         & 46.61           \\
VDI~\cite{VDI:conf/cvpr/LuoHGJL23}                     & CVPR'23               &                        & -              & 52.32          & 31.37          & -              & -              & 48.09          & 28.76         & -               \\
UniVTG~\cite{UniVTG:conf/iccv/LinZCPGWYS23}                  & ICCV'23               &                        & 72.63          & 60.19          & 38.55          & 52.17          & -              & -              & -             & -               \\ 
\hline
VCA~\cite{VCA:conf/mm/WangCJ21}                     & MM'21                 & \multirow{4}{*}{WS}    & 58.58          & 38.13          & 19.57          & 38.49          & 50.45          & 31.00          & -             & 33.15           \\
CNM~\cite{CNM:conf/aaai/ZhengHCL22}                     & AAAI'22               &                        & 60.04          & 35.15          & 14.95          & 38.11          & 55.68          & 33.33          & 13.29         & 37.55           \\
CPL~\cite{CPL:conf/cvpr/ZhengHCPL22}                     & CVPR'22               &                        & 65.99          & 49.05          & 22.61          & 43.23          & 55.73          & 31.37          & 13.68         & 36.65           \\
Huang et al.~\cite{Huang:conf/cvpr/HuangYS23}            & CVPR'23               &                        & 69.16          & 52.18          & 23.94          & 45.20          & 58.07          & 36.91          & -             & 41.02           \\ 
\hline
PSVL~\cite{PSVL:conf/iccv/NamAKHC21}                    & ICCV'21               & \multirow{5}{*}{US}    & 46.47          & 31.29          & 14.17          & 31.24          & 44.74          & 30.08          & 14.74         & 29.62           \\
Gao et al.~\cite{Gao:journals/tcsv/GaoX22}              & TCSVT'21              &                        & 46.69          & 20.14          & 8.27           & -              & 46.15          & 26.38          & 11.64         & -               \\
DSCNet~\cite{DSCNet:conf/aaai/LiuQWDZCXZ22}                  & AAAI'22               &                        & 44.15          & 28.73          & 14.67          & -              & 47.29          & 28.16          & -             & -               \\
PZVMR~\cite{PZVMR:conf/mm/WangWLY22}                   & MM'22                 &                        & 46.83          & 33.21          & 18.51          & 32.62          & 45.73          & 31.26          & 17.84         & 30.35           \\
Kim et al.~\cite{Kim:conf/wacv/KimPLPS23}              & WACV'23               &                        & 52.95          & 37.24          & 19.33          & 36.05          & 47.61          & 32.59          & 15.42         & 31.85           \\ 
\hline
LLM4VG~\cite{LLM4VG:journals/corr/abs-2312-14206} & ArXiv'23 & \multirow{3}{*}{ZS}  & 25.97 & 10.91 & 3.47 & -           & -     & -              & -             & -    \\
UniVTG*~\cite{UniVTG:conf/iccv/LinZCPGWYS23}              & ICCV'23               &    & 44.09          & 25.22          & 10.03          & 27.12          & -              & -              & -             & -               \\
Ours                    & -                     &                        & \textbf{52.69} & \textbf{33.01} & \textbf{15.89} & \textbf{34.87} & \textbf{40.67} & \textbf{22.49} & \textbf{9.42} & \textbf{27.21}  \\
\bottomrule
\end{tabular}}}
\caption{Performance comparison of methods under different setups on the test set of Charades-STA~\cite{CharadesSTA:conf/iccv/GaoSYN17} and ActivityNet-Captions~\cite{ActivityNetCaptions:conf/iccv/KrishnaHRFN17}. FS denotes fully supervised, WS denotes weakly supervised, US denotes unsupervised, and ZS denotes zero-shot. UniVTG* represents the results reported in UniVTG~\cite{UniVTG:conf/iccv/LinZCPGWYS23} under the zero-shot setting. }
\vspace{-3mm}
\label{tab:sota}
\end{table*}

\subsection{Datasets}
To evaluate the effectiveness of our approach, we perform experiments on three widely used datasets: Charades-STA, ActivityNet Captions, and TACoS. Our method stands out as it operates without any training, refraining from the use of training data during the experiments, with results being directly derived and calculated on the test sets.

\noindent {\textbf{Charades-STA}} \cite{CharadesSTA:conf/iccv/GaoSYN17} The work presented by Gao et al. \cite{CharadesSTA:conf/iccv/GaoSYN17} builds upon the Charades dataset \cite{Charades:conf/eccv/SigurdssonVWFLG16} for action recognition and localization. They tailored the Charades dataset for Video Moment Retrieval (VMR) by incorporating query annotations. The resulting Charades-STA dataset comprises 6,670 videos with 16,124 associated queries. The average duration of the videos is 30.59 seconds, while the moments have an average duration of 8.09 seconds. Among the 16,124 queries, there are 37 long moments (\(L_{\text{moment}}/L_{\text{vid}} \geq 0.5\)) included in this dataset.

\noindent {\textbf{ActivityNet-Captions}} 
\cite{ActivityNetCaptions:conf/iccv/KrishnaHRFN17}  is compiled specifically for the task of video captioning, sourced from ActivityNet \cite{ActivityNet:conf/cvpr/HeilbronEGN15}, where videos are annotated with 200 distinct activity classes. This dataset, named ActivityNet-Captions, comprises 19,811 videos accompanied by 71,957 queries. On average, the videos span approximately 117.75 seconds, with individual moments lasting around 37.14 seconds. Within this dataset, there are 15,736 instances of long moments out of the total 71,957 queries.

\noindent {\textbf{TACoS}} \cite{TACoS:journals/tacl/RegneriRWTSP13} comprises 127 videos sourced from MPIICooking \cite{MPIICooking:conf/eccv/RohrbachRAAPS12}. It encompasses a total of 18,818 video-text pairs depicting cooking activities in the kitchen, meticulously annotated by Regneri et al. \cite{TACoS:journals/tacl/RegneriRWTSP13}.

\subsection{Evaluation Metric}
To assess the effectiveness of our model, we employ well-established metrics, including R@tIoU (Recall at temporal Intersection over Union) and mIoU (mean Intersection over Union), as delineated in previous research \cite{2DTAN:conf/aaai/ZhangPFL20, MMN:conf/aaai/00010WLW22, VDI:conf/cvpr/LuoHGJL23}, to ensure a comprehensive and fair evaluation. Specifically, we measure temporal alignment accuracy by calculating the temporal Intersection over Union (tIoU) between predicted and ground-truth boundaries. R@tIoU quantifies the proportion of predictions exceeding predefined thresholds, namely \{0.3, 0.5, 0.7\}. Conversely, mIoU offers a consolidated evaluation by averaging the IoU scores across all predictions, providing a holistic perspective of model performance across various thresholds.

\subsection{Implementation Details}
In the Instruction-Refined Video Captioning stage, we primarily utilize VideoChatGPT~\cite{VideoChatGPT:journals/corr/abs-2306-05424} as the Video LLM, and conduct ablation experiments on VideoChat2~\cite{VideoChat2:journals/corr/abs-2311-17005} as well. We adhere to the acceptable input frame numbers for each Video LLM during the video retrieval process. For the Query-caption Matching stage, Sentence-BERT~\cite{} is employed as the text feature extraction model, and cosine similarity calculation is used to evaluate the correlation between queries and captions. The initial number of segments is set to 5, and the refinement sliding window has a width of 10 with a step size of 5. These settings are thoroughly compared and explained in the ablation experiments. Our experiments don't involve any training processes, and inference is the only requirement. All experiments are conducted on an NVIDIA A100 GPU.

\begin{table}
\centering
\scalebox{0.8}{
\setlength{\tabcolsep}{2mm}{
\begin{tabular}{c|c|cccc} 
\toprule
Method      & Setup               & R@0.3         & R@0.5         & R@0.7         & mIoU           \\ 
\hline
2D TAN~\cite{2DTAN:conf/aaai/ZhangPFL20}      & \multirow{4}{*}{FS} & 40.01         & 27.99         & 12.92         & 27.22          \\
VSLNet~\cite{VSLNet:conf/acl/ZhangSJZ20}      &                     & 35.54         & 23.54         & 13.15         & 24.99          \\
MDETR~\cite{MDETR:conf/iccv/KamathSLSMC21}       &                     & 37.97         & 24.67         & 11.97         & 25.49          \\
UniVTG~\cite{UniVTG:conf/iccv/LinZCPGWYS23}      &                     & 56.11         & 43.44         & 24.27         & 38.63          \\ 
\hline
UniVTG*~\cite{UniVTG:conf/iccv/LinZCPGWYS23}   & \multirow{2}{*}{ZS} & 5.17          & 1.27          & 0.27          & 4.40           \\
Ours        &                     & \textbf{8.12} & \textbf{3.74} & \textbf{1.25} & \textbf{5.54}  \\
\bottomrule
\end{tabular}}}
\caption{Performance comparison of methods under different setups on the test set of TACoS~\cite{TACoS:journals/tacl/RegneriRWTSP13}. FS denotes fully supervised and ZS denotes zero-shot. UniVTG* represents the results reported in UniVTG~\cite{UniVTG:conf/iccv/LinZCPGWYS23} under the zero-shot setting. }
\vspace{-3mm}
\label{tab:tacos}
\end{table}

\subsection{Comparison with Existing Methods}
We compare our ChatVTG with previous methods across fully supervised, weakly supervised, unsupervised, and zero-shot settings.
The performance on Charades-STA and ActivityNet-Captions is reported in Tab.~\ref{tab:sota}.
On Charades-STA, our ChatVTG significantly outperforms previous zero-shot methods with an improvement of 8.6\% R@0.3 (44.09\%$\rightarrow$52.69\%).
For ActivityNet-Captions, the performance approaches the previous unsupervised methods, indicating the robustness of our ChatVTG.
Additionally, we report the results on TACoS in Tab.~\ref{tab:tacos}. The results indicate that our method outperforms zero-shot UniVTG.

\subsection{Ablation Studies}
To showcase the individual efficacy of every module within our ChatVTG framework, we conduct comprehensive ablation studies on the Charades-STA dataset.

\noindent {\textbf{Impact of Different Instructions.}}  We first investigate the impact of using different instructions. As shown in \cref{tab:instruction}, 
the performance comparison from captions generated under different instructions yields relatively similar outcomes. 
This suggests that queries cover a range of perspectives, with each query potentially focusing on different aspects. 
However, a notable deviation is observed in the ``Dressing" instruction, indicating that certain scenarios may heavily or entirely lack key descriptions related to attire. 
Depending solely on dressing may not provide sufficient accuracy in event timing matching. 
Conversely, instructions such as ``Action," ``Place," ``Emotion," and ``Interaction" are frequently utilized to differentiate and describe specific segments more effectively.

\noindent {\textbf{Different Matching Computation.}} 
In our exploration of effectively leveraging multi-granularity captions to match queries and obtain the final moment accurately, we have experimented with various computation methods for merging similarity scores of multi-granularity captions. Reference to \cref{matrix}, computation methods in~\cref{tab:fusion} are as follows:

\noindent{(1) Baseline: The baseline method involves calculating the cosine similarity score between the captions generated solely from the ``action" granularity and the query.}

\noindent{(2) Normalization After Summation: 
This method adds up the cosine similarity score of all granularities before normalizing.
With the incorporated information from multiple granularities, it makes a more comprehensive understanding of the video content.
However, potential redundancy or conflicting information between different granularities may lead to performance fallback in certain scenarios.
}
\begin{table}
\centering
\scalebox{0.8}{
\setlength{\tabcolsep}{3mm}{
\begin{tabular}{c|cccc} 
\toprule
Instruction & mIoU & R@0.3 & R@0.5 & R@0.7   \\ 
\hline
Action      & 32.73 & 51.05 & 29.54 & 13.90  \\
Place       & 32.46 & 50.48 & 27.37 & 13.60  \\
Dressing    & 27.38 & 42.83 & 21.82 & 9.10   \\
Emotion     & 31.71 & 48.47 & 28.66 & 14.49  \\
Interaction & 31.15 & 48.22 & 27.13 & 13.61  \\
\bottomrule
\end{tabular}}}
  \caption{Experimental results using different instructions.}
  \label{tab:instruction}
  \vspace{-3mm}
\end{table}

\noindent{(3) Summation after Normalization: 
This method normalizes the cosine similarity score of all granularities before adding up.
Therefore, each granularity contributes equally to the overall similarity score, regardless of its inherent scale. 
However, similar to method~(2), it may be sensitive to outliers or uneven distributions within the data.}

\noindent{(4) Normalization after Row-wise Maximum: 
This method identifies the granularity with the highest correlation with the given query and considers its score as the overall caption score.
However, this approach overlooks the correlations between the given query and other granularity captions and may lead to biased results.}

\noindent{(5) Normalization after Column-wise Maximum: 
This method selects the maximum similarity score for each column (video clip) and then normalizes these scores. 
It comprehensively considers the maximum potential similarity between each granularity caption and the query while allowing for a fair comparison across different granularity levels. 
It effectively captures the diverse perspectives provided by each granularity level and results in improved performance, as adopted in our final implementation.}

In summary, given the rationale and performance of the ``Normalization after Column-wise Maximum" approach, we opt to utilize this method for matching computations.

\begin{table}
\centering
\scalebox{0.8}{
\setlength{\tabcolsep}{3mm}{
\begin{tabular}{c|cccc} 
\toprule
Fusion method & mIoU & R@0.3 & R@0.5 & R@0.7   \\ 
\hline
(1)           & 32.73 & 51.05 & 29.54 & 13.90  \\
(2)           & 34.17 & 51.71 & 31.62 & 14.85  \\
(3)           & 34.43 & 52.45 & 31.67 & 15.32  \\
(4)           & 33.30 & 51.48 & 31.75 & 15.24  \\
(5)           & \textbf{34.69} & \textbf{52.77} & \textbf{32.90} & \textbf{15.70}  \\
\bottomrule
\end{tabular}}}
\vspace{-2mm}
  \caption{Experimental results of selecting the final moment through different matching computation methods.}
  \vspace{-1mm}
  \label{tab:fusion}
\end{table}

\noindent {\textbf{Impact of Clip Number.}} To understand the impact of the number of clips, we conducted ablation experiments on the coarsely segmented clips. We tested different numbers of clips, specifically 3, 5, 10, and 20. As shown in~\cref{tab:clip_num}, the results indicate that the performance is relatively better when the number of clips is 5. Therefore, we have selected 5 as the initial number of coarsely segmented clips.

\begin{table}
\centering
\scalebox{0.8}{
\setlength{\tabcolsep}{3mm}{
\begin{tabular}{c|cccc} 
\toprule
Clip Num. & mIoU  & R@0.3 & R@0.5 & R@0.7   \\ 
\hline
3         & 32.28 & 49.56 & 29.78 & 12.55  \\
5         & \textbf{32.42} & \textbf{49.98} & \textbf{29.67} & \textbf{13.39}  \\
10        & 30.96 & 47.79 & 28.47 & 12.41  \\
20        & 28.19 & 42.50 & 25.11 & 11.10  \\
\bottomrule
\end{tabular}}}
\vspace{-2mm}
  \caption{Experimental results of different clip numbers. }
  \vspace{-1mm}
  \label{tab:clip_num}
\end{table}

\noindent {\textbf{Impact of Slide Window Size.}} 
To achieve more precise moments, we partition the coarse-grained moments into smaller and denser time windows using a sliding window approach, followed by re-captioning the segments. 
The baseline performance represents the no sliding window used method, as shown in~\cref{tab:slide_window}. 
We denote the window width and step size for the sliding window as (wide, step). 
The results show that employing a sliding window improves performance, especially when the window is set to (10, 5). 
However, adopting a sliding window for the entire duration increases computational complexity significantly.
Additionally, it is not meaningful to refine segments with low matching scores. 
Therefore, we only select coarse-grained moments with an initial IoU greater than 0.7 for refinement, with a window setting of (10, 5). 
The results of this approach, as indicated as the ``Coarse-to-Fine", show a performance improvement compared to the baseline.

\begin{table}
\centering
\scalebox{0.8}{
\setlength{\tabcolsep}{3mm}{
\begin{tabular}{c|cccc} 
\toprule
Slide window   & mIoU          & R@0.3          & R@0.5          & R@0.7            \\ 
\hline
Baseline       & 32.73          & 51.05          & 29.54          & 13.90           \\
(20, 5)        & 32.54          & 51.02          & 31.34          & 12.90           \\
(10, 5)        & 34.09          & 52.69          & 34.19          & 14.92           \\
(10, 2)        & 32.56          & 50.94          & 31.67          & 13.60           \\
Coarse-to-Fine & \textbf{34.87} & \textbf{52.79} & \textbf{33.04} & \textbf{15.89}  \\
\bottomrule
\end{tabular}}}
\vspace{-2mm}
  \caption{Experimental results of different slide window settings.}
  \vspace{-3mm}
  \label{tab:slide_window}
\end{table}

\noindent {\textbf{Performance of Different Video LLMs}}  Testing is also conducted on another Video LLM, VideoChat~\cite{VideoChat2:journals/corr/abs-2311-17005}, which exhibited superior video understanding performance compared to VideoChatGPT. As shown in~\cref{tab:videollm}, an enhancement in the captioning capability of Video LLMs leads to an overall improvement in VTG performance.

\subsection{Qualitative Results}

\begin{figure}[t!]
  \centering
    \includegraphics[width=0.9\linewidth]{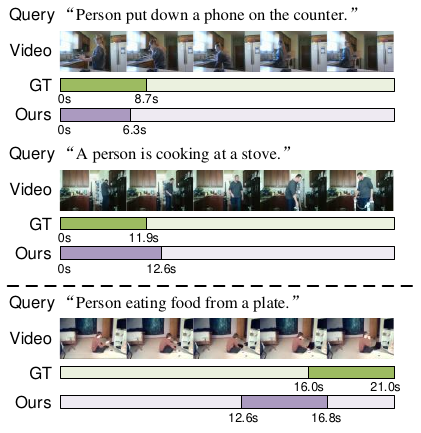}
    \vspace{-3mm}
  \caption{Above the dashed line are successful examples on the Charades-STA dataset, and below are the failed ones.
  }
  \vspace{-3mm}
  \label{fig:visualize}
\end{figure}

Fig.~\ref{fig:visualize} displays selected prediction outcomes. Successful predictions from the Charades-STA dataset appear above the dashed line, while unsuccessful ones are presented below. Our observations indicate that predictions tend to be more precise for videos featuring simpler, prolonged actions. In contrast, accuracy diminishes in videos characterized by intricate action sequences and rapid alternation of multiple actions.

\begin{table}
\centering
\scalebox{0.8}{
\begin{tabular}{c|cccc} 
\toprule
Video LLM     & R@0.3 & R@0.5 & R@0.7 & mIoU   \\ 
\hline
Video-ChatGPT & 34.87 & 52.79 & 33.04 & 15.89  \\
VideoChat2    & 36.55 & 56.56 & 34.52 & 16.34  \\
\bottomrule
\end{tabular}}
\vspace{-2mm}
  \caption{Experimental results of different Video LLMs}
  \vspace{-3mm}
  \label{tab:videollm}
\end{table}

\section{Conclusion and Future Work}
\label{sec:conclusion}

We introduce ChatVTG in this paper, a novel and effective approach for zero-shot VTG that leverages the advanced capabilities of Video Dialogue LLMs. 
The success of ChatVTG signifies a leap forward in VTG tasks, offering a scalable and efficient solution for understanding and interacting with video content through natural language. 
In the future, our work can be applied to online video search, as its ability to enable the pre-processing of offline conversion from videos to textual descriptions. 
It can also be used for cross-video matching and retrieving, beyond the confines of individual videos.

\textbf{Acknowledgements.} This work was supported by the Fundamental Research Funds for the Central Universities (2023JBZD003), the National NSF of China (No.U23A20314), the National Key R\&D Program of China under Grant No. 2021ZD0112100.

\newpage
{
    \small
    \bibliographystyle{ieeenat_fullname}
    \bibliography{main}
}


\end{document}